\documentclass[pra,aps,twocolumn,amsmath,amssymb,superscriptaddress]{revtex4-1}
\usepackage{epsfig,amsmath}
\usepackage{subfigure}
\usepackage{graphicx}
\usepackage{dcolumn}
\usepackage{stmaryrd}
\usepackage{mathrsfs}
\usepackage{pifont}
\usepackage{amsthm}
\usepackage{amssymb}
\usepackage{bm}
\usepackage{latexsym}
\usepackage[colorlinks=true,linkcolor=cyan,citecolor=cyan]{hyperref}
\usepackage{color}
\usepackage{epstopdf}
\usepackage[ruled]{algorithm2e}

\begin{document}

\title{Generic eigenstate preparation via measurement-based purification}

\author{Jia-shun Yan}
\affiliation{School of Physics, Zhejiang University, Hangzhou 310027, Zhejiang, China}

\author{Jun Jing}
\email{Email address: jingjun@zju.edu.cn}
\affiliation{School of Physics, Zhejiang University, Hangzhou 310027, Zhejiang, China}

\date{\today}

\begin{abstract}
It is not a general opinion that that a quantum system could be purified into a target eigenstate via repeated measurements on a coupled qubit rather than direct transitions in the Hamiltonian. The projective measurement on the ancillary qubit gives rise to the positive operator-valued measures on the system that can filter out the unwanted states except the target one. In application, we discuss the measurement-based entanglement purification by which maximally entangled states (Bell states and Greenberger-Horne-Zeilinger states) can be distilled from the maximally mixed states or separable states. We also demonstrate the significant acceleration of a stimulated Raman adiabatic passage assisted by similar measurements. Our scheme allows arbitrary eigenstate preparation and reveals efficiency in multipartite systems for subspace purification. It offers a promising and generic quantum-control framework enriching the functionalities of quantum measurement.
\end{abstract}
\maketitle

\section{Introduction}

Quantum state preparation is a basic and crucial premise for plenty of modern quantum applications, including but not limited to measurement-based quantum computation~\cite{OneWay,MeasurementComputation}, quantum teleportation~\cite{QuantumTeleportation,QuantumTeleportation2}, quantum dense coding~\cite{DenseCoding}, and quantum cryptography~\cite{Cryptography}. Preparing eigenstates, especially for a complex system, is of great importance in quantum chemistry~\cite{FermionicHamEigenstates,QuantumChemistry} and condensed-matter physics~\cite{ManybodyLocalization,ManybodyRMP}. Various interesting tools have been applied in pushing the system of interest into a target eigenstate, including entanglement generation by dissipation~\cite{Dissipative,PreparationByMarkovProcess}, variational quantum algorithms~\cite{VQA,VQE}, and shortcuts to adiabaticity~\cite{STAThreelevel,STA,STAQi}. Among them, distilling a mixed state into a desired pure state of a high fidelity distinguishes itself since any quantum system is inevitably coupled to an external environment. It is therefore reasonable to find that state purification and entanglement purification have developed as key technologies in quantum information and quantum computation~\cite{PurificationViaNoisyChannels,MixedStateEntanglement,QubitPurification,PurificationZhong}.

Frequent quantum measurements over a noncommutative operator with respect to the Hamiltonian could freeze the measured quantum system at an eigenstate by asymptotically affecting the system dynamics, known as the quantum Zeno effect~\cite{ZenoParadox}. When the measurement operator becomes parametric dependent, the measured system could be steered to a target state from either a pure state~\cite{AharonovZeno,NonselectiveAharonovZeno} or a mixed state~\cite{MeasurementEvolution,NonselectiveMeasurementEvolution} through a finite number of measurements with nonvanishing measurement intervals. As the dimension of the system becomes larger, however, it is more difficult to perform direct measurements on the system. Quantum engineering could be alternatively realized through indirect measurements on an ancillary system. In general, a projective measurement or postselection on the ancillary system gives rise to a positive operator-valued measure (POVM) on the interested system~\cite{MeasurementCharging,Naimark}, which can be navigated to a target state with a finite probability~\cite{MeasurementPurification}. The indirect measurement method has a wide range of applications associated with state purification, such as cooling a resonator to its ground state~\cite{MeasurementCooling,MeasurementCoolingPRL,ExternalLevelCooling,ExpOneModeCooling,HeraldedControlMotion}, enhancing the bath spin polarization~\cite{SpinBathState,Polarization}, and charging a quantum battery~\cite{NonselevtiveCharging,MeasurementCharging}. Nevertheless, the purification of the system with repeated measurements on the ancillary system is under certain constraints, e.g., the target states cannot be arbitrarily chosen~\cite{MeasurementPurification}, the target system is required to be nondegenerate~\cite{MeasurementEigensolver}, and all the system eigenstates have to be connected directly or indirectly through given transitions~\cite{MeasurementSteering}. It is then desired to find a generic scheme capable of distilling a degenerate or nondegenerate system into an eigenstate with a limited number of ancillary systems.

In this work, we propose a general scheme that an interested system can be purified into an arbitrary eigenstate by repeatedly measuring a coupled ancillary qubit. Rather than directly transferring the system population to the target state, we use projective measurements to filter out the populations on the other states, which can be shuffled through a purification operator. By virtue of the population renormalization in comply with the non-unitary operations determined by measurements, there would be a unique population rise on the target state. Our scheme is applied to entanglement purification by generating the Bell states and the Greenberger-Horne-Zeilinger (GHZ) states. It exemplifies a creation of maximally entangled states~\cite{SingletPreparation,QuantumEntanglement} from maximally mixed states or a separable state. To prepare the Bell state, a recent scheme based on the nonselective measurements~\cite{MeasurementSteering} assigns  one detector (ancillary qubit) for each transition channel towards the target state and relies on three-body interactions. In sharp contrast, our scheme requires only a single ancillary qubit and is efficient in operation. It can be applied to the GHZ state preparation and shows potential to avoid many-body interactions in the subspace purification. Also our framework of state purification by measurement can be integrated with the standard stimulated Raman adiabatic passage (STIRAP), by which a complete population transfer is promoted even in a diabatic passage.

The rest part of this work is structured as follows. In Sec.~\ref{ProbPurification}, a generic framework is introduced for eigenstate preparation based on repeated measurements on the ancillary qubit. A necessary condition for state purification is established through defining a purification operator in the system space, given knowledge about the energy spectrum of the system. In Sec.~\ref{Uniqueness}, we show that the steady state of the system in the limit of an infinite number of measurements is exactly the same as the target state with a given purification operator. In Sec.~\ref{UniquenessTotal}, our framework is extended to a more general system Hamiltonian and a necessary condition is provided about the eigenstate preparation. In Sec.~\ref{Sec:PrepareEntangledState}, we apply our framework to prepare a singlet Bell state for a double-qubit system and the GHZ state for a three-qubit system. In Sec.~\ref{Sec:STIRAP}, we present two hybrid models combining STIRAP and state purification to demonstrate an accelerated adiabatic passage in a three-level system. We summarize our work in Sec.~\ref{Sec:Conclusion}.

\section{Theoretical framework}\label{Sec:Model}

In general, two necessary conditions have to be fulfilled for state purification through quantum measurements. The first condition is that the population over the target state would be ever increased upon a desired measurement outcome until approaching unit. The second condition is that the target state is the unique one approached by the system under a sufficient number of measurements. In Secs.~\ref{ProbPurification} and \ref{Uniqueness}, we illustrate our scheme on accumulating the population over the target state and the coincidence between the steady state and the target state when the system Hamiltonian in the interaction picture is time-independent and in a specific formation. Then in Sec.~\ref{UniquenessTotal}, the system Hamiltonian is relieved to a general form, by which we discuss the purification condition through measurements.

\subsection{Purification operator and probabilistic purification}\label{ProbPurification}

\begin{figure}[htbp]
\centering
\includegraphics[width=0.95\linewidth]{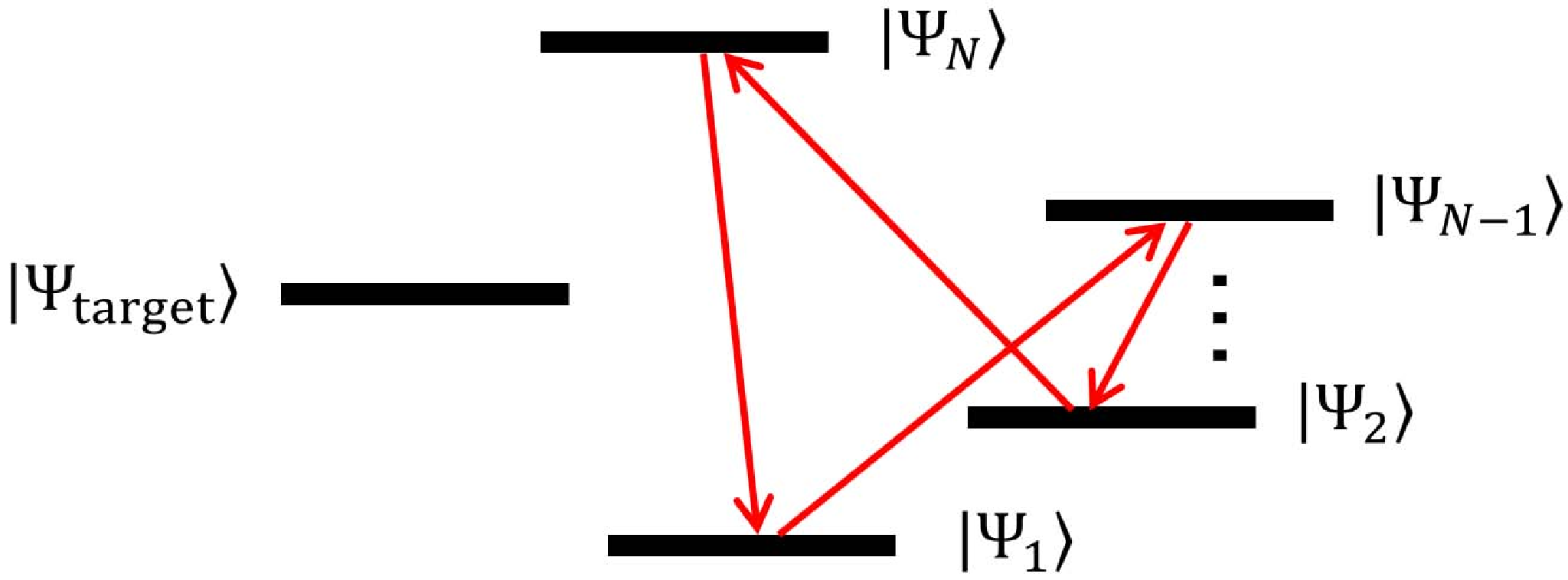}
\caption{Transition diagram of the purification operator $Q$ in an $N+1$ dimensional system, which can be arbitrarily designed provided all the eigenstates of the system except the target one are directly or indirectly connected with no isolated element. $Q$ does not hold a directional path towards the target eigenstate and allows self-transition (projection) operators of the unwanted eigenstates. }\label{Fig:Qoperator}
\end{figure}

In our framework of state preparation and purification by measurement, the target state $|\Psi_{\rm target}\rangle$ is a given eigenstate of the system Hamiltonian $H_S$, i.e., $H_S|\Psi_{\rm target}\rangle=\lambda|\Psi_{\rm target}\rangle$, where $\lambda$ is the eigenvalue. We have an ancillary qubit with a free Hamiltonian $H_A$. With a purification operator $Q$ built up in Fig.~\ref{Fig:Qoperator}, we have a purification Hamiltonian
\begin{equation}\label{Ham}
H_P=g_a\left(A^\dagger Q+AQ^\dagger\right)=g_a\left[\begin{array}{cc}0&Q \\ Q^\dagger&0 \end{array}\right].
\end{equation}
Here $g_a$ is the coupling strength between the system and the ancillary qubit. $A\equiv|\varphi\rangle\langle\varphi_\perp|$ and $A^\dagger\equiv|\varphi_\perp\rangle\langle\varphi|$ are transition operators for the ancillary qubit about the initial state $|\varphi\rangle$ and its orthogonal counterpart $|\varphi_\perp\rangle$, i.e., $\langle\varphi|\varphi_{\perp}\rangle=0$. $H_P$ is also the interaction Hamiltonian in the interaction picture with respect to $H_0=H_S+H_A$. To illustrate the underlying mechanism of our scheme and show the uniqueness of the purified state, $H_P$ is assumed to be time independent for simplicity. It is valid if the target system is resonant with the ancillary qubit. We discuss in Sec.~\ref{UniquenessTotal} the general situation using the full Hamiltonian in the Schr\"odinger picture.

To purify the system into the target state, the system operator $Q$ is constructed by connecting all the unwanted eigenstates of the system $H_S$. The target state is required to be a dark state of $Q$, i.e.,
\begin{equation}\label{PurificationCondition}
Q|\Psi_{\rm target}\rangle=0.
\end{equation}
It means that $Q$ forbids the transitions from $|\Psi_{\rm target}\rangle$ to the other eigenstates. Yet it does not forbid the inverse transitions. In another word, when $Q\neq Q^\dagger$, either $Q^\dagger|\Psi_{\rm target}\rangle\neq0$ or $Q^\dagger|\Psi_{\rm target}\rangle=0$ has no bearing on our scheme. Figure~\ref{Fig:Qoperator} is an instance of $Q^\dagger|\Psi_{\rm target}\rangle=0$, where no transition channel connects the target eigenstate.

Given the Hamiltonian in Eq.~(\ref{Ham}), the joint time-evolution operator of both system and ancillary qubit for a period of $\tau$ is
\begin{equation}\label{Utot}
U(\tau)=\left[\begin{array}{cc}C^T(\tau)&S^\dagger(\tau)\\S(\tau)&C(\tau)\end{array}\right],
\end{equation}
where the Kraus operators are
\begin{equation}\label{KrausOperator}
\begin{aligned}
&C(\tau)=\sum_{k=0}^N\frac{(-i\tau)^{2k}}{(2k)!}\left(Q^\dagger Q\right)^k,\\
&S(\tau)=\sum_{k=0}^N\frac{(-i\tau)^{2k+1}}{(2k+1)!}\left(Q^\dagger Q\right)^kQ^\dagger,
\end{aligned}
\end{equation}
respectively. The time evolution of the whole system under the Hamiltonian $H_P$ is repeatedly interrupted by the instantaneous projective measurement about the initial state of the ancillary system $M_\varphi\equiv|\varphi\rangle\langle\varphi|$. Any two neighboring measurements can be arbitrarily spaced in time, so that our method is essentially robust against the systematic error about the measurement moments. After $m-1$ rounds of evolution and successful measurement, the whole system state is $\rho_{\rm tot}=\rho_s(t)\otimes|\varphi\rangle\langle\varphi|$, where $t=\sum_{j=1}^{m-1}\tau_j$ with $\tau_j$'s indicating the measurement intervals. Then the whole system state becomes
\begin{equation}\label{rhotot}
\begin{aligned}
\rho_{\rm tot}(t+\tau_m)=&\frac{M_\varphi U(\tau_m)\rho_s\otimes|\varphi\rangle\langle\varphi|U^\dagger(\tau_m)M_\varphi}{P_\varphi^{(m)}}\\
=&\rho_s(t+\tau_m)\otimes|\varphi\rangle\langle\varphi|
\end{aligned}
\end{equation}
after one more round of evolution and measurement lasting $\tau_m$, where $P_\varphi^{(m)}\equiv{\rm Tr}[C(\tau_m)\rho_s(t)C^\dagger(\tau_m)]$ represents the success probability of the $m$th round. In particular, if the measurement outcome is as desired, i.e., the ancillary qubit is reset as the initial state, then the system and the ancillary qubit are decoupled from each other as in Eq.~(\ref{rhotot}). According to the Naimark's dilation theorem~\cite{Naimark}, the projective measurements performed on the ancillary qubit induce a POVM $\mathcal{M}(\tau)[\mathcal{O}]\equiv C(\tau)\mathcal{O}C^\dagger(\tau)$ acting on the system. Then the system state in Eq.~(\ref{rhotot}) can be expressed as
\begin{equation}\label{rhos}
\rho_s(t+\tau_m)=\frac{C(\tau_m)\rho_s(t)C^\dagger(\tau_m)}{P_\varphi^{(m)}},
\end{equation}
according to the time-evolution operator in Eq.~(\ref{Utot}).

Since the purification operator annihilates the target state $Q|\Psi_{\rm target}\rangle=0$, we have $C(\tau_m)|\Psi_{\rm target}\rangle=|\Psi_{\rm target}\rangle$ due to the first line in Eq.~(\ref{KrausOperator}), i.e., the POVM $\mathcal{M}(\tau_m)$ does not immediately change the population over the target state
\begin{equation}
\langle\Psi_{\rm target}|\mathcal{M}(\tau_m)[\rho_s(t)]|\Psi_{\rm target}\rangle=\langle\Psi_{\rm target}|\rho_s(t)|\Psi_{\rm target}\rangle.
\end{equation}
As $C(\tau)$ is a non-unitary operator, the system state should be renormalized by the measurement probability $P_\varphi^{(m)}$. Then $1/P_\varphi\geq1$ in either Eq.~(\ref{rhotot}) or Eq.~(\ref{rhos}) acts as a gain factor raising the target-state population. It follows with a purification inequality:
\begin{equation}
\langle\Psi_{\rm target}|\rho_s(t+\tau_m)|\Psi_{\rm target}\rangle\geq\langle\Psi_{\rm target}|\rho_s(t)|\Psi_{\rm target}\rangle.
\end{equation}
A successful measurement then suggests that the ancillary qubit in its initial state heralds a population rise over the target state.

After $m$ rounds of measurements, the overlap between the system state and the target state $F_m\equiv\langle\Psi_{\rm target}|\rho_s(\sum_{j=1}^m\tau_j)|\Psi_{\rm target}\rangle$, i.e., the target-state fidelity, could be expressed as
\begin{equation}\label{fidelity}
F_m=\frac{F_{m-1}}{P_\varphi^{(m)}}=\frac{F_{m-2}}{P_\varphi^{(m)}P_\varphi^{(m-1)}}=\frac{\langle\Psi_{\rm target}|\rho_s(0)|\Psi_{\rm target}\rangle}{\prod_{j=1}^{m}P_\varphi^{(j)}}.
\end{equation}
Therefore the system can be gradually purified under more and more rounds of evolution and measurement, provided that its initial population over the target state is not vanishing. Nevertheless, the outcome of our framework is probabilistic as indicated by the denominator of the last equivalence in Eq.~(\ref{fidelity}), i.e., the success probability $P_s=\prod_{j=1}^{m}P_\varphi^{(j)}$. One can find that $P_s$ is lower-bounded by the initial population over the target state.

\subsection{Uniqueness of purified state}\label{Uniqueness}

Under a sufficiently large number $m$ of measurements, the measurement probability of the ensued rounds would approach unit $P_\varphi^{(j)}\rightarrow1$, $j\geq m$. Otherwise the target-state fidelity will keep growing with no upper bound. In this situation, the ancillary qubit is freezed at the initial state $|\varphi\rangle$, the system approaches a steady state, and the success probability $P_s$ becomes invariant in time. We can show that the steady state is exactly the target state $|\Psi_{\rm target}\rangle$.

Note that the Kraus operator $C(\tau)$ in Eq.~(\ref{KrausOperator}) is Hermitian and $C(\tau_m)|\Psi_{\rm target}\rangle=|\Psi_{\rm target}\rangle$. Then the operator can always be expanded as
\begin{equation}
\begin{aligned}
C(\tau_m)&=|\Psi_{\rm target}\rangle\langle\Psi_{\rm target}|\\ &+ {\sum_k}'\epsilon_k(\tau_m)|\psi_k(\tau_m)\rangle\langle\psi_k(\tau_m)|,
\end{aligned}
\end{equation}
where $|\psi_k(\tau_m)\rangle$'s and $\epsilon_k(\tau_m)$'s are instantaneous eigenstates and eigenvalues of $C(\tau_m)$, respectively. ${\sum_k}'$ indicates the summation over all degrees of freedom except the target state, whose eigenvalue is one. Consequently, the measurement probability of the $m$th round can be written as
\begin{equation}\label{UnitProb}
\begin{aligned}
&P_\varphi^{(m)}={\rm Tr}\left[C(\tau_m)\rho_s(t)C^\dagger(\tau_m)\right] \\
&=F_{m-1}+{\sum_k}'\epsilon_k^2(\tau_m)\langle\psi_k(\tau_m)|\rho_s\left(\sum_{j=1}^{m-1}\tau_j\right)|\psi_k(\tau_m)\rangle.
\end{aligned}
\end{equation}
The second part on the right hand side of Eq.~(\ref{UnitProb}) is associated with $\tau_m$-dependent populations on the instantaneous eigenstates. And $\tau_m$ can be randomly chosen in our framework. A $\tau_m$-independent and close-to-unit measurement probability $P_\varphi^{(j\geq m)}\rightarrow1$ therefore requires the vanishing populations over $|\psi_k(\tau_m)\rangle$, suggesting that the system has been successfully prepared at the target state $\rho_s(t)=|\Psi_{\rm target}\rangle\langle\Psi_{\rm target}|$.

Alternatively, one can find that the square of the eigenvalues of the Kraus operator $C(\tau_m)$ are always lower than or equivalent to one. In particular, we have
\begin{equation}
\begin{aligned}
\epsilon_k^2(\tau_m)&=\left|C(\tau_m)|\psi_k(\tau_m)\rangle\right|^2
=\left|\langle\varphi|U(\tau_m)|\varphi\rangle|\psi_k(\tau_m)\rangle\right|^2\\
&\leq\left|U(\tau_m)|\varphi\rangle|\psi_k(\tau_m)\rangle\right|^2=1.
\end{aligned}
\end{equation}
It means that most populations on these eigenstates are reduced by measurements, except the target state and some special states satisfying $\epsilon_k^2(\tau_m)=1$ for the $m$th measurement performed at the moment $t+\tau_m$. However, since the measurement intervals for the free joint evolutions can be randomly chosen, the protection over the population of such unwanted states cannot last for a sufficient number of evolution-measurement rounds. Only the target-state population will eventually survive.

Our state-purification framework therefore promises the uniqueness of the purified state, as long as the population on the target state is not vanishing at the initial time. The POVM induced by the purification operator acts as a sieve to filter out the populations on unwanted states.

\subsection{Purification by general Hamiltonian}\label{UniquenessTotal}

Under certain conditions, our purification framework can be generalized to accommodate the effective Hamiltonian beyond the compact form in Eq.~(\ref{Ham}). In the Schr\"odinger picture, we can consider the full Hamiltonian as
\begin{equation}\label{fullHam}
H=H_0+H_P=H_S+H_A+H_P,
\end{equation}
which consists of the system Hamiltonian, the ancillary-qubit Hamiltonian, and the purification Hamiltonian expressed by Eq.~(\ref{Ham}). For simplicity, $H$ is assumed to be time independent. $H_A=\omega_aA^{\dagger}A$ with $\omega_a$ the characteristic frequency of the ancillary qubit. With the same projective measurement $M_{\varphi}$ on the initial state of the ancillary qubit, the nonunitary operator for a period of free evolution can be expanded as
\begin{equation}\label{Vtau}
\begin{aligned}
V(\tau)&\equiv\langle\varphi|e^{-iH\tau}|\varphi\rangle=\sum_n\frac{(-i\tau)^n}{n!}\langle\varphi|H^n|\varphi\rangle\\
&=\sum_n\frac{(-i\tau)^n}{n!}V^{(n)},
\end{aligned}
\end{equation}
where $V^{(n)}\equiv\langle\varphi|H^n|\varphi\rangle$. Note $V(\tau)$ reduces to $C(\tau)$ if $H$ can be written as $H_P$ in the interaction picture. Recalling the purification Hamiltonian $H_P=g_a(A^\dagger Q+AQ^\dagger)$, we have $\langle\varphi| H_P|\varphi\rangle=0$. Then only the items consisting of an even number of $H_P$ can survive in expansion. The first three orders of $V^{(n)}$ are
\begin{equation}
\begin{aligned}
&V^{(1)}=\langle\varphi|H|\varphi\rangle=H_S,\\
&V^{(2)}=\langle\varphi|H^2|\varphi\rangle=H_S^2 + g_aQ^\dagger Q,\\
&V^{(3)}=\langle\varphi|H^3|\varphi\rangle=H_S^3\\
&+g_a^2(H_S Q^\dagger Q + \omega_a Q^\dagger Q+Q^{\dagger} Q H_S+Q^{\dagger} H_S Q).
\end{aligned}
\end{equation}

One can find that $Q^\dagger$ and $Q$ appear by ordered pairs in each order of $V^{(n)}$. In particular, $V^{(n)}=H_S^n+D_Q^{(n)}$, where $H_S^n$ is the system Hamiltonian to the $n$th power and $D_Q^{(n)}=D_Q^{(n)}(H_S, Q^\dagger, Q)$ is a function of $H_S$ and ordered pairs of $Q^\dagger$ and $Q$. According to the definition about the purification operator in Eq.~(\ref{PurificationCondition}), we have $D_Q^{(n)}|\Psi_{\rm target}\rangle=0$ and $D_Q^{(n)}|\Psi_k\rangle={\sum_l}'\beta_{lk}^{(n)}|\Psi_l\rangle$, where ${\sum_l}'$ represents the summation over all eigenstates of $H_S$ except the target state and $\beta_{lk}^{(n)}$ is the overlap coefficient between $|\Psi_k\rangle$ and $|\Psi_l\rangle$ under $D_Q^{(n)}$.

Then the POVM induced by measuring the initial state of the ancillary qubit $|\varphi\rangle$ gives rise to the system state
\begin{equation}
\begin{aligned}
&\rho_s(t+\tau)\sim\mathcal{M}[\rho_s(t)]=\sum_{n,m}\frac{(-i\tau)^n(i\tau)^m}{n!m!}V^{(n)}\rho_s(t)V^{(m)}\\
&=\sum_{n,m}\frac{(-i\tau)^n(i\tau)^m}{n!m!}\left[H_S^n+D_Q^{(n)}\right]\rho_s(t)\left[H_S^m+D_Q^{(m)}\right].
\end{aligned}
\end{equation}
The population over each eigenstate of the target system reads
\begin{equation}\label{population}
\begin{aligned}
&\langle\Psi_k|\mathcal{M}[\rho_s(t)]|\Psi_k\rangle=\rho_{kk}(t)+{\sum_l}'\bigg[e^{i\lambda_k\tau}\alpha_{lk}^*(\tau)\rho_{lk}(t)\\
&+e^{-i\lambda_k\tau}\alpha_{lk}(\tau)\rho_{kl}(t)+{\sum_j}'\alpha_{jk}^*(\tau)\alpha_{lk}(\tau)\rho_{jl}(t)\bigg],
\end{aligned}
\end{equation}
where $\rho_{ij}(t)\equiv\langle\Psi_i|\rho_s(t)|\Psi_j\rangle$ and $\alpha_{lk}(\tau)\equiv\sum_n\frac{(i\tau)^n}{n!}\beta_{lk}^{(n)}$. The measurement probability of the current round can be obtained by summing over Eq.~(\ref{population}) for every eigenstate
\begin{equation}
P_\varphi=\sum_k\langle\Psi_k|\mathcal{M}[\rho_s(t)]|\Psi_k\rangle=1+\chi(\tau),
\end{equation}
where
\begin{equation*}
\begin{aligned}
\chi(\tau)=&{\sum_{l,k}}'\bigg[e^{i\lambda_k\tau}\alpha_{lk}^*(\tau)\rho_{lk}(t)
+e^{-i\lambda_k\tau}\alpha_{lk}(\tau)\rho_{kl}(t)\\
+&{\sum_j}'\alpha_{jk}^*(\tau)\alpha_{lk}(\tau)\rho_{jl}(t)\bigg]
\end{aligned}
\end{equation*}
is a function of the density-matrix elements in the subspace orthogonal to the target state. Generally, we have $-1\leq \chi(\tau)\leq0$. As discussed in Sec.~\ref{Uniqueness}, $P_\varphi$ is close to unit and the target system approaches a steady state after a sufficient number of measurements. In this situation, $\chi(\tau)\rightarrow0$. Assuming that the magnitude of the coherent elements is negligible in comparison to the population when $\rho_s(t)$ becomes invariant with time, then $\chi(\tau)\rightarrow0$ renders
\begin{equation*}
\begin{aligned}
{\sum_k}'\bigg[e^{i\lambda_k\tau}\alpha_{kk}^*(\tau)\rho_{kk}(t)&+e^{-i\lambda_k\tau}\alpha_{kk}(\tau)\rho_{kk}(t)\\
&+{\sum_{l}}'|\alpha_{lk}(\tau)|^2\rho_{ll}(t)\bigg]=0.
\end{aligned}
\end{equation*}
Since it is independent of $\tau$, then the populations on the unwanted eigenstates have to be vanishing: ${\sum_k}'\rho_{kk}(t)=0$. It also indicates that the steady state under measurements is coincident with the target one.

\section{Preparation of entanglement}\label{Sec:PrepareEntangledState}

Quantum information processing often requires entangled states as a resource for, e.g., the working qubits in the quantum teleportation protocol. However, a quantum system with channels to the external environment would be in general ended with a mixed state~\cite{PurificationViaNoisyChannels,MixedStateEntanglement}. It is thus interesting and important to distill or generate entangled states from such a quantum system. In this section, we apply our state-purification scheme to prepare Bell states and GHZ state from the maximally mixed states $\rho_s(0)=I/d$~\cite{MixedState} where $d$ is the dimension of the system and $I$ is the identity matrix, which means the initial system population has an even distribution over all the eigenstates. Moreover for the GHZ state, we introduce an efficient purification operator that avoids the many-body interactions. And our scheme is shown to be robust in the presence of nonideal purification operator.

\subsection{Bell state preparation}\label{Sec:BellState}

\begin{figure}[htbp]
\centering
\includegraphics[width=1\linewidth]{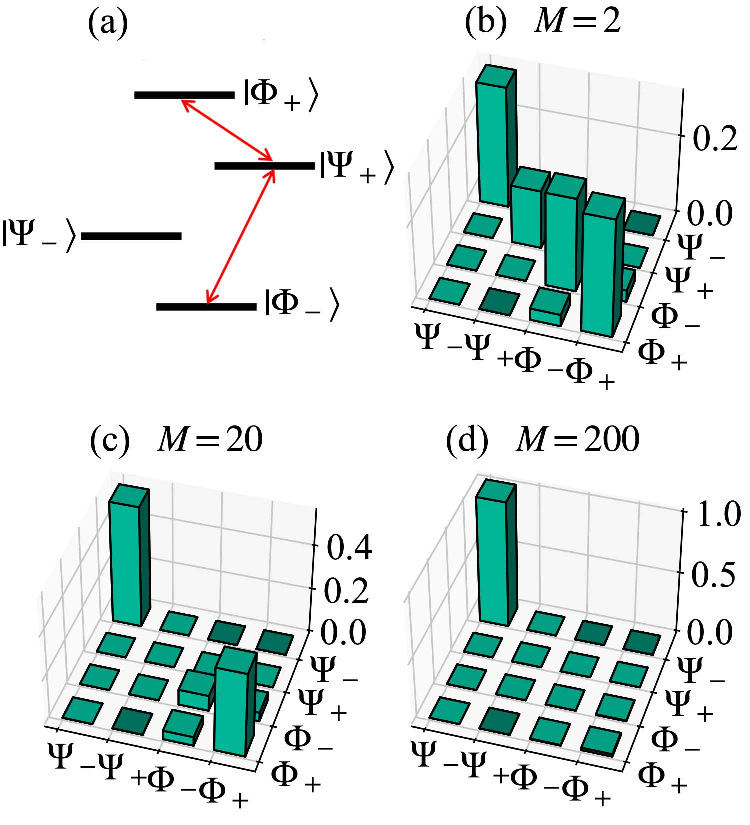}
\caption{(a) Transition diagram for the purification operator $Q$ defined in Eq.~(\ref{Q_BellState}), where the bidirectional arrows represent the back-and-forth transitions between a pair of eigenstates. (b)-(d) Tomographies of density matrix in the system eigenbasis after (b) $M=2$, (c) $M=20$, and (d) $M=200$ rounds of measurements. The measurement interval for each round is $\tau_i=\tau_0+\delta t$, where $\delta t$ is a uniformly distributed random number in the region $(-\tau_0/2, \tau_0/2)$. $\tau_0=2/\omega_0$, $|\omega_a|=\omega_0$, $g_a/\omega_0=0.2$, and $g_s=5g_a$.}\label{Fig:BellStates}
\end{figure}

We first choose the singlet Bell state $|\Psi_{\rm target}\rangle=|\Psi_-\rangle=(|01\rangle-|10\rangle)/\sqrt{2}$ as the target stat, where $|0\rangle$ and $|1\rangle$ represent the excited state $|e\rangle$ and the ground state $|g\rangle$ of the target qubits, respectively. For a double-qubit system with a strong $XX$ coupling~\cite{XXCoupling}, the system Hamiltonian can be written as ($\hbar=1$)
\begin{equation}\label{SystemHam}
H_S=\omega_1\sigma_1^+\sigma_1^-+\omega_2\sigma_2^+\sigma_2^-+g_s\sigma_1^x\sigma_2^x,
\end{equation}
where $\omega_i$ and $\sigma_i^{\pm}$ are respectively the bare frequency and the transition operators for the $i$th qubit and $g_s$ represents the coupling strength between qubit-$1$ and qubit-$2$. Under the resonant condition, $\omega_1=\omega_2=\omega_0$, the target state is one of the system eigenstate $H_S|\Psi_{\rm target}\rangle=(\omega_0-g_s)|\Psi_{\rm target}\rangle$. To purify the system into the target state, the purification Hamiltonian can be chosen as
\begin{equation}
H_P=g_a\left(A^\dagger Q+AQ^\dagger\right), \quad Q=\sigma_1^++\sigma_2^+,
\end{equation}
which can be straightforwardly verified to satisfy the condition $Q|\Psi_{\rm target}\rangle=0$. The initial state of the ancillary qubit is set as the excited state $|\varphi\rangle=|e\rangle$. Thus $A=\sigma_a^+=|e\rangle\langle g|$, $A^\dagger=\sigma_a^-=|g\rangle\langle e|$, and the ancillary-qubit Hamiltonian is $H_A=\omega_a\sigma_a^-\sigma_a^+$. The system qubits and the ancillary qubit are resonant and their coupling strength is assumed to be much weaker than the interaction between the two system qubits, i.e., $g_a\ll g_s$. It is then reasonable to neglect the counter-rotating terms $\sigma_a^+\sigma_i^+$ and $\sigma_a^-\sigma_i^-$ in $H_P$ under the rotating-wave approximation. Then the full Hamiltonian $H=H_S+H_A+H_P$ reads
\begin{equation}
\begin{aligned}
H=&\sum_{i=1,2}\omega_i\sigma_i^+\sigma_i^- + \omega_a\sigma_a^-\sigma_a^++g_s\sigma_1^x\sigma_2^x\\
&+g_a\left[\sigma_a^-(\sigma_1^++\sigma_2^+)+\sigma_a^+(\sigma_1^-+\sigma_2^-) \right].
\end{aligned}
\end{equation}

The rest three eigenstates of the system Hamiltonian in Eq.~(\ref{SystemHam}) are
\begin{equation}\label{BellEigenstates}
\begin{aligned}
|\Psi_+\rangle=&\frac{1}{\sqrt{2}}(|01\rangle+|10\rangle), \\
|\Phi_-\rangle=&\frac{1}{\xi_-}\left[\left(\sqrt{g_s^2+\omega_0^2}-\omega_0\right)|00\rangle-g_s|11\rangle\right], \\
|\Phi_+\rangle=&\frac{1}{\xi_+}\left[\left(\sqrt{g_s^2+\omega_0^2}+\omega_0\right)|00\rangle+g_s|11\rangle\right],
\end{aligned}
\end{equation}
where $\xi_\pm\equiv\sqrt{g_s^2+(\omega_0\pm\sqrt{g_s^2+\omega_0^2})^2}$ are the normalization coefficients. And their eigenvalues are $\omega_0+g_s$, $\omega_0-\sqrt{\omega_0^2+g_s^2}$, and $\omega_0+\sqrt{\omega_0^2+g_s^2}$, respectively. In the eigenbasis of the system, the purification operator can be rewritten as
\begin{equation}\label{Q_BellState}
\begin{aligned}
Q=&\frac{\xi_+}{\sqrt{2(g_s^2+\omega_0^2)}}|\Phi_+\rangle\langle\Psi_+|
+\frac{\xi_-}{\sqrt{2(g_s^2+\omega_0^2)}}|\Phi_-\rangle\langle\Psi_+|\\
&+\frac{\sqrt{2}g_s}{\xi_+}|\Psi_+\rangle\langle\Phi_+|-\frac{\sqrt{2}g_s}{\xi_-}|\Psi_+\rangle\langle\Phi_-|,
\end{aligned}
\end{equation}
which involves with the transitions among $|\Phi_+\rangle$, $|\Phi_-\rangle$, and $|\Psi_+\rangle$ as shown in Fig.~\ref{Fig:BellStates}(a). The coefficient for each transition in the operator $Q$ indicates the variation rate of the eigenstates' population except the target one.

The system states after a certain number of rounds of evolution and projective measurement $M_{\varphi}$ with random measurement intervals are demonstrated in Fig.~\ref{Fig:BellStates}(b)-(d). The initial state is the maximally mixed state with population equally distributed on all eigenstates, which can be remarkably modified by only two rounds of measurements [see Fig.~\ref{Fig:BellStates}(b)]. After $M=20$ measurements [see Fig.~\ref{Fig:BellStates}(c)], the population over the target state $|\Psi_-\rangle$ has been raised from $0.25$ to $0.55$ (over one half); the population over $|\Psi_+\rangle$ becomes almost vanishing; and the populations over $|\Phi_-\rangle$ and $|\Phi_+\rangle$ are about $0.07$ and $0.38$, respectively. These results can be understood by the transition rates presented in $Q$ operator in Eq.~(\ref{Q_BellState}). For example, the transition rate for $|\Phi_+\rangle\rightarrow|\Psi_+\rangle$ is $\sqrt{2}g_s/\xi_+$, whose magnitude is the smallest one among all the rates, leading to an inefficient population transfer from $|\Phi_+\rangle$ to the other states. After $M=200$ measurements [see Fig.~\ref{Fig:BellStates}(d)], all the populations have been cumulated onto the target state, which means the other states are filtered out by the measurement-induced purification. In particular, it is found that the state fidelity is $F\approx0.98$ and the success probability is $P_s\approx26\%$.

\begin{figure}[htbp]
\centering
\includegraphics[width=0.95\linewidth]{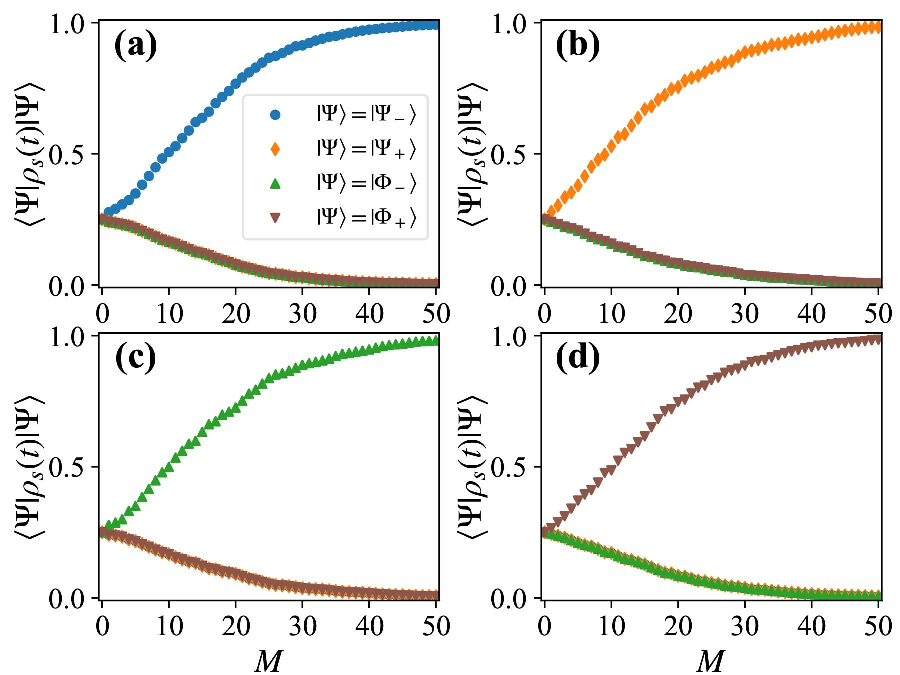}
\caption{Population dynamics of eigenstates as functions of the measurement number with various purification operators: (a) $Q_{\Psi_-}$, (b) $Q_{\Psi_+}$, (c) $Q_{\Phi_-}$, and (d) $Q_{\Phi_+}$. The other parameters are the same as those in Fig.~\ref{Fig:BellStates}.}\label{Fig:AllEigenStates}
\end{figure}

Our framework of measurement-based purification is much simpler than the state steering based on the nonselective measurements~\cite{MeasurementSteering}, which employs one ancillary qubit for every transition from unwanted states to the target state and also involves with the three-body interactions. Alternatively, the number of ancillary qubits can be reduced at the cost of extra unitary rotations with respect to the system Hamiltonian, which requires frequently switching on and off the interaction between the system and the ancillary qubit. In contrast, our framework contains only a single ancillary qubit and two-body interactions, because of no direct transitions to the target state.

The target state is arbitrary in our framework. Based on the basic condition in Eq.~(\ref{PurificationCondition}), three main recipes can be followed to design the purification operator for a desired target state: (i) $Q=\sum_k'a_k|\Psi_{\rm target}\rangle\langle\Psi_k|$, collecting the transitions from all the other eigenstates to the target state; (ii) $Q=\sum_k'a_k|\Psi_k\rangle\langle\Psi_{k+1}|$, building transitions between every pair of neighboring states ordered in a certain way, e.g., the annihilation operator of a resonator; (iii) $Q=\sum_k'a_k|\Psi_k\rangle\langle\Psi_k|$, mapping all the other eigenstates to themselves, i.e., a collection of projective operators. Here $a_k$'s are arbitrary and nonvanishing coefficients.

For the two-qubit system in Eq.~(\ref{SystemHam}), all the eigenstates in Eq.~(\ref{BellEigenstates}) could be prepared via the measurement-based purification. The purification performance as well as the purification operators could be efficiently simulated in digital quantum circuits~\cite{51QubitsMeasurementVQE}. We here follow the third recipe in the preceding discussion to demonstrate the purification process, which avoids setting up transitions among unwanted states and those towards the target state. And for simplicity, we suppose all the self-projectors are the same in weight. Then the purification operators for the eigenstates in Eq.~(\ref{BellEigenstates}) could be constructed as
\begin{equation}\label{fourQ}
Q_{\Psi} = I-|\Psi\rangle\langle\Psi|,\quad\Psi\in\{\Psi_-,\Psi_+,\Phi_-,\Phi_+\}.
\end{equation}

Figure~\ref{Fig:AllEigenStates} demonstrates the population dynamics about the four eigenstates of Hamiltonian~(\ref{SystemHam}) with various target states, whose purification operators could be given by Eq.~(\ref{fourQ}). It is found that the system can be prepared as the desired target eigenstates within $M=50$ rounds of free evolution and measurement. They are ended with a fidelity over $F=0.98$ and a success probability over $P_s=25\%$. The populations over the unwanted states decrease gradually to zero with almost the same rate. The general purification operator in Eq.~(\ref{fourQ}) can involve with many-body interactions, when the target state is a multi-particle entangled state or the system eigenstructure becomes degenerate. In the following section, it is shown that our framework could be still efficient if initially the target state is the only occupied one in its degenerate subspace.

\subsection{GHZ state preparation}\label{Subsec:GHZstate}

\begin{figure}[htbp]
\centering
\includegraphics[width=1\linewidth]{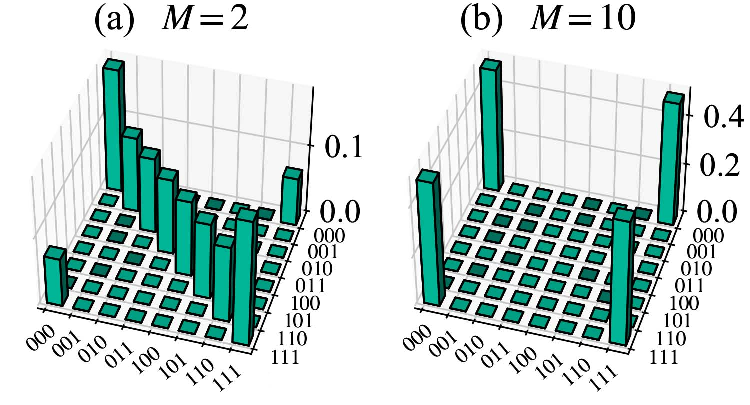}
\caption{Tomographies of density matrix for the Ising chain of three qubits after (a) $M=2$ and (b) $M=10$ rounds of evolution and measurement. The system state starts from a mixed state with an even distribution over populations on system eigenstates. The measurement interval for each round is $\tau_i=\tau_0+\delta t$, where $\delta t$ is a uniformly distributed random number in the region $(-\tau_0/2, \tau_0/2)$. $\tau_0=2/J$, $g_a/J=0.4$, and $\omega_a/J=1$.}\label{Fig:GHZState}
\end{figure}

Our measurement-based purification framework can adapt to generating entangled state for multiple qubit system. In the absence of a transversal magnetic field, we consider a one-dimensional chain of three spins-$1/2$ linked by the nearest-neighbor Ising bonds, i.e., $H_S=J(\sigma_1^z\sigma_2^z+\sigma_2^z\sigma_3^z)$, where $J$ is the coupling strength~\cite{IsingModel}. It is a typical degenerate system whose eigenbasis is not unique determined. The target state is set as one of the maximally entangled state for this discrete system:
\begin{equation}\label{GHZ}
|{\rm GHZ}\rangle=|\Psi_{1,+}\rangle=\frac{1}{\sqrt{2}}(|000\rangle+|111\rangle).
\end{equation}
And the other eigenstates of the system can also be regarded as the general GHZ states:
\begin{equation}\label{GHZeigenstates}
\begin{aligned}
&|\Psi_{1,-}\rangle=\frac{1}{\sqrt{2}}(|000\rangle-|111\rangle),\\
&|\Psi_{2,\pm}\rangle=\frac{1}{\sqrt{2}}(|010\rangle\pm|101\rangle),\\
&|\Psi_{3,\pm}\rangle=\frac{1}{\sqrt{2}}(|001\rangle\pm|100\rangle),\\
&|\Psi_{4,\pm}\rangle=\frac{1}{\sqrt{2}}(|011\rangle\pm|110\rangle).
\end{aligned}
\end{equation}

To filter out the populations on these unwanted eigenstates, the purification operator can be constructed by a collection of self-projectors as in Eq.~(\ref{fourQ}):
\begin{equation}\label{GHZQ}
\begin{aligned}
Q&=I-|\Psi_{1,+}\rangle\langle\Psi_{1,+}|\\
&=|\Psi_{1,-}\rangle\langle\Psi_{1,-}| + |\Psi_{2,+}\rangle\langle\Psi_{2,+}| + |\Psi_{2,-}\rangle\langle\Psi_{2,-}|\\
& + |\Psi_{3,+}\rangle\langle\Psi_{3,+}|+|\Psi_{3,-}\rangle\langle\Psi_{3,-}| + |\Psi_{4,+}\rangle\langle\Psi_{4,+}|\\
& + |\Psi_{4,-}\rangle\langle\Psi_{4,-}|.
\end{aligned}
\end{equation}
For this system, the effective Hamiltonian can be the purification Hamiltonian in the formation of Eq.~(\ref{Ham}), where $A^\dagger=\sigma^+=|e\rangle\langle g|$. The system states after $M=2$ and $M=10$ rounds of measurements are shown in Fig~\ref{Fig:GHZState}(a) and Fig~\ref{Fig:GHZState}(b), respectively. It is found that even from the maximally mixed state, the system can be purified into the valuable GHZ state by several random measurements. After $M=10$ measurements, the state fidelity is close to unit and the success probability is $P_s=12.5\%$, which is equivalent to the initial population on the target state.

For both Bell state and GHZ state, the initial state of the interested system so far is chosen as the maximally mixed state with the maximal von Neumann entropy $S[\rho_s(0)]=\log(d)$~\cite{VonNeumann}. It is surely a ``hard mode'' choice for state purification. In a less extreme condition, e.g., when the system starts from a state with vanishing population on the eigenstate that is degenerate with the target state, the many-body interactions in Eq.~(\ref{GHZQ}) could be avoided and then our scheme could become scalable. As for discriminating the unwanted degenerate state, the techniques beyond our scheme, e.g., the parity subspace projections~\cite{PurityProjectionNN,PurityProjectionPRL}, could be applied to deal with a more general initial state.

\begin{figure}[htbp]
\centering
\includegraphics[width=1\linewidth]{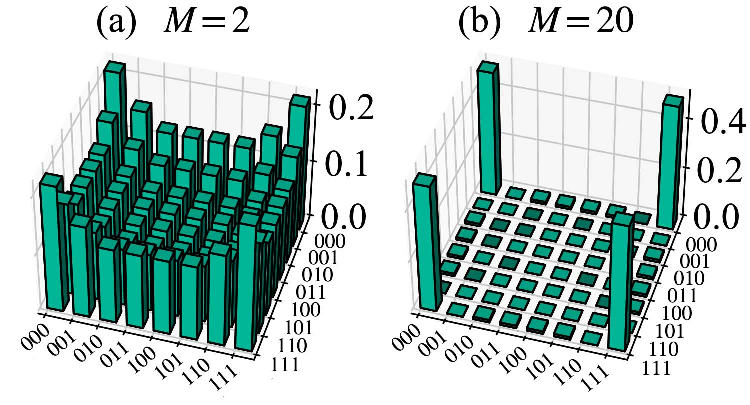}
\caption{Tomographies of density matrix for the system after (a) $M=2$ and (b) $M=20$ rounds of evolution and measurement under $Q'$. The system state starts from a superposed state $|+\rangle|+\rangle|+\rangle$. $g_a/J=0.2$, $\omega_a/J=1$, and $\tau_0=2/J$.}\label{Fig:GHZState2}
\end{figure}

We here introduce an alternative purification operator based on subspace purification, which is able to purify the system from a separable state into the GHZ state with only two-body interactions between the ancillary qubit and the target qubits. Following the necessary condition in Eq.~(\ref{PurificationCondition}), it could be designed as $Q'=a\sigma_1^z+b\sigma_2^z-(a+b)\sigma_3^z$ with $ab\neq0$, and
\begin{equation}
\begin{aligned}
&Q'|\Psi_{1,\pm}\rangle=0,\\
&Q'|\Psi_{2,\pm}\rangle=-2b|\Psi_{2,\mp}\rangle,\\
&Q'|\Psi_{3,\pm}\rangle=2a|\Psi_{3,\mp}\rangle+b(|\Psi_{3,\mp}\rangle+|\Psi_{3,\pm}\rangle),\\
&Q'|\Psi_{4,\pm}\rangle=2a|\Psi_{4,\mp}\rangle+b(|\Psi_{4,\mp}\rangle-|\Psi_{4,\pm}\rangle).
\end{aligned}
\end{equation}
In measurement-based purification, $Q'$ can be used to filter out all the populations outside the degenerate space spanned by $|\Psi_{1,\pm}\rangle$. If the system state is initially orthogonal to the unwanted degenerate state $|\Psi_{1,-}\rangle$, then the target GHZ state $|\Psi_{1,+}\rangle$ could be eventually attained by a sufficient number of measurements. In this case, the purification Hamiltonian in Eq.~(\ref{Ham}) becomes
\begin{equation}
H_P=g_a\sigma_a^x(\sigma_1^z+\sigma_2^z-2\sigma_3^z),
\end{equation}
where $\sigma_a^x=A+A^{\dagger}$ with $A=\sigma_a^-$. It involves only two-body ZX interactions that can be realized in superconducting qubits~\cite{ZXcouplingPRB,ZXcouplingPRL}. In Fig.~\ref{Fig:GHZState2}, we demonstrate the system tomographies during the purification by measurements, where the initial state is set as $|+\rangle|+\rangle|+\rangle$ with $|+\rangle=(|e\rangle+|g\rangle)/\sqrt{2}$. The system is transformed from a state that is diversely populated in the whole space to the GHZ state. After $M=20$ measurements, the populations over all unwanted states are almost filtered out and the final fidelity of the target GHZ state approaches unit.

\begin{figure}[htbp]
\centering
\includegraphics[width=1\linewidth]{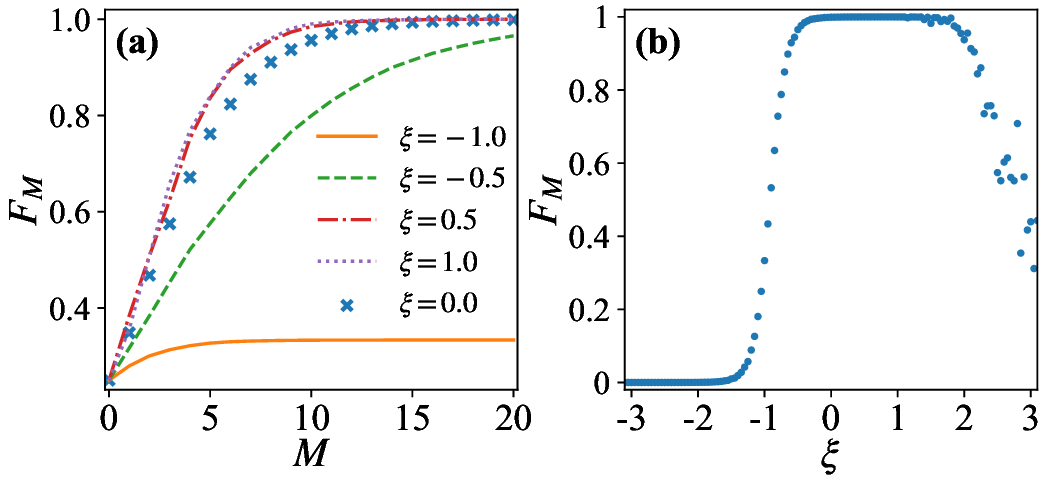}
\caption{(a) Fidelities of the GHZ state as a function of measurement number with errors in $Q'$. (b) Fidelities of the GHZ state after $M=20$ measurements as a function of the error. Parameters are the same as those in Fig.~\ref{Fig:GHZState2}.}\label{IdealNonideal}
\end{figure}

The efficiency of our scheme can be benchmarked by evaluating the fidelity under rounds of evolution and measurement in the presence of imperfections in purification operator. In practice, it is assumed that
\begin{equation}
Q'=\sigma_1^z+\sigma_2^z-(2+\xi)\sigma_3^z,
\end{equation}
where $\xi$ represents an error in the nonideal purification operator. In Fig.~\ref{IdealNonideal}, the fidelities of the GHZ state $F_M=\langle\Psi_{1,+}|\rho_s(t=\sum_{j=1}^M\tau_j)|\Psi_{1,+}\rangle$ are plotted as a function of the measurement number $M$ and the value of $\xi$. It is interesting to find that an asymmetry dependence of the fidelity on $\xi$. As shown in Fig.~\ref{IdealNonideal}(a), the purification process is slightly accelerated under a positive $\xi$ and the final results of $\xi=0.5$ and $\xi=1.0$ are almost the same as the ideal case $\xi=0$. In contrast, the final fidelity of $M=20$ significantly declines under a negative $\xi$. For $\xi=-1.0$, we have $F_M\approx0.33$. Figure~\ref{IdealNonideal}(b) presents that in a wide range of error, about $[-0.5, 2]$, the fidelity can be maintained above $0.9$. This result justifies the robustness of our scheme in realization of the purification operator.

\section{Acceleration of adiabatic passage}\label{Sec:STIRAP}

\begin{figure}[htbp]
\centering
\includegraphics[width=1\linewidth]{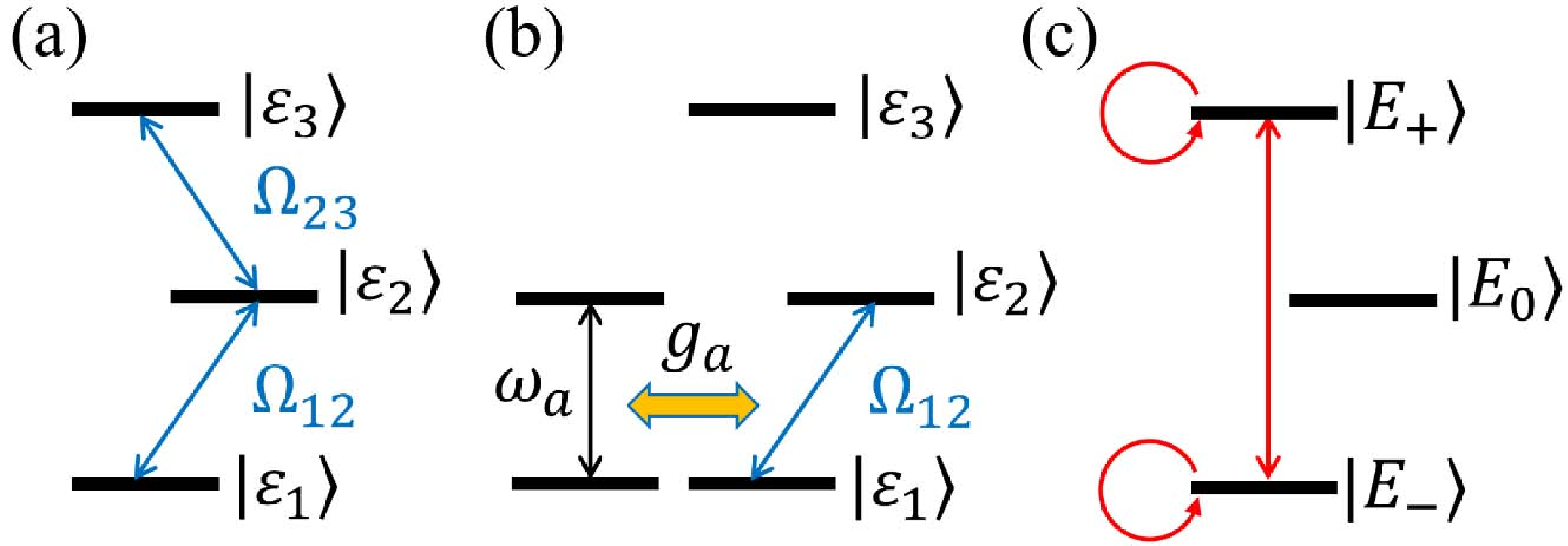}
\caption{(a) Sketch for the conventional STIRAP protocol in a three-level cascade system, where two resonant driving fields $\Omega_{12}$ and $\Omega_{23}$ are coupled to the transitions of $|\varepsilon_1\rangle\leftrightarrow|\varepsilon_2\rangle$ and $|\varepsilon_2\rangle\leftrightarrow|\varepsilon_3\rangle$, respectively. (b) State-purification model for an ancillary qubit of frequency $\omega_a$ coupled to the three-level system with a coupling strength $g_a$. To push the system into the pure state $|\varepsilon_3\rangle$, the driving field $\Omega_{23}$ is removed due to the setting of our state-purification scheme in Fig.~\ref{Fig:Qoperator}. (c) Transition diagram of the purification operator $Q$ provided by Eq.~(\ref{Q_STIRAP}) in the eigenbasis of the system. Directed cycles represent the self-transitions for the unwanted eigenstates. }\label{Fig:StirapModel}
\end{figure}

The conventional stimulated Raman adiabatic passage in a three-level system is used to faithfully transfer the population on an eigenstate to another one with dark states~\cite{STAThreelevel}. It could be realized by properly driving the transitions in the system as shown in Fig.~\ref{Fig:StirapModel}(a). Under the assumption that the external driving fields are resonant with the corresponding frequency splittings between the driven levels, the system Hamiltonian in the interaction picture can be written as~\cite{AdiabaticBattery}
\begin{equation}\label{StirapHam}
H_1 = \Omega_{12}(t)\sigma_{12}^x+\Omega_{23}(t)\sigma_{23}^x,
\end{equation}
where $\sigma_{12}^x\equiv|\varepsilon_1\rangle\langle\varepsilon_2|+|\varepsilon_2\rangle\langle\varepsilon_1|$ and $\sigma_{23}^x\equiv|\varepsilon_2\rangle\langle\varepsilon_3|+|\varepsilon_3\rangle\langle\varepsilon_2|$ are the transition operators in the three-level system. One of eigenstates of the system
\begin{equation}
|E_0(t)\rangle=\frac{1}{\sqrt{2}}
\left[\frac{\Omega_{23}(t)}{\Delta(t)}|\varepsilon_1\rangle-\frac{\Omega_{12}(t)}{\Delta(t)}|\varepsilon_3\rangle\right]
\end{equation}
constitutes the time-dependent adiabatic path for state engineering. In particular, when the system is initialized at the ground state $|\varepsilon_1\rangle$, a perfect population transfer to $|\varepsilon_3\rangle$ could be realized by a slowly-decreasing field $\Omega_{23}(t)$ and a slowly-increasing field $\Omega_{12}(t)$. If the Rabi frequencies of these fields are rapidly varying with time, then the system evolution can deviate significantly from the adiabatic path, resulting in an incomplete population transfer to $|\varepsilon_3\rangle$. Our state-purification scheme could be applied to modify, complete, and accelerate STIRAP.

In our theoretical framework, STIRAP is integrated with a purification Hamiltonian consisting of only one resonant driving field $\Omega_{12}(t)$ for the chosen target state $|\Psi_{\rm target}\rangle=|\varepsilon_3\rangle$, i.e., the system Hamiltonian now reads
\begin{equation}\label{HSstirap}
H_S= \Omega_{12}(t)\sigma_{12}^x.
\end{equation}
In accordance with state purification, the target state is an eigenstate of the system Hamiltonian $H_S|\varepsilon_3\rangle=0$. As shown in Fig.~\ref{Fig:StirapModel}(b), the purification operator associated with the ancillary qubit can be set as $Q=\sigma_{12}^-=|\varepsilon_1\rangle\langle\varepsilon_2|$. Straightforwardly one can confirm the purification condition $Q|\varepsilon_3\rangle=0$. The initial state of the qubit is the ground state $|\varphi\rangle=|g\rangle$ that determines $A$. Then the purification Hamiltonian is
\begin{equation}
H_P=g_a\left(\sigma_a^+\sigma_{12}^-+\sigma_a^-\sigma_{12}^+\right)
\end{equation}
where $\sigma_{12}^+=|\varepsilon_2\rangle\langle\varepsilon_1|$. The full Hamiltonian of our scheme in the rotating frame with respect to $H_0=\sum_{i=1}^3\varepsilon_i|\varepsilon_i\rangle\langle\varepsilon_i|+\omega_a\sigma_a^+\sigma_a^-$ reads
\begin{equation}
H=H_S+H_P=\Omega_{12}(t)\sigma_{12}^x+g_a\left(\sigma_a^+\sigma_{12}^-+\sigma_a^-\sigma_{12}^+\right),
\end{equation}
where the ancillary qubit is assumed to be resonant with the splitting between $|\varepsilon_1\rangle$ and $|\varepsilon_2\rangle$, i.e., $\omega_a=\varepsilon_2-\varepsilon_1$. For the system Hamiltonian in Eq.~(\ref{HSstirap}), the three eigenstates are
\begin{equation}
\begin{aligned}
|E_0\rangle&=|\varepsilon_3\rangle, \\
|E_-\rangle&=\frac{1}{\sqrt{2}}(|\varepsilon_1\rangle-|\varepsilon_2\rangle), \\
|E_+\rangle&=\frac{1}{\sqrt{2}}(|\varepsilon_1\rangle+|\varepsilon_2\rangle),
\end{aligned}
\end{equation}
by which the purification operator $Q$ could be rewritten as
\begin{equation}\label{Q_STIRAP}
Q=|E_-\rangle\langle E_+|-|E_+\rangle\langle E_-|+|E_+\rangle\langle E_+|-|E_-\rangle\langle E_-|.
\end{equation}
Transitions presented in the operator $Q$ are demonstrated in Fig.~\ref{Fig:StirapModel}(c), where there is no transition towards the target state $|\varepsilon_3\rangle$.

\begin{figure}[htbp]
\centering
\includegraphics[width=0.95\linewidth]{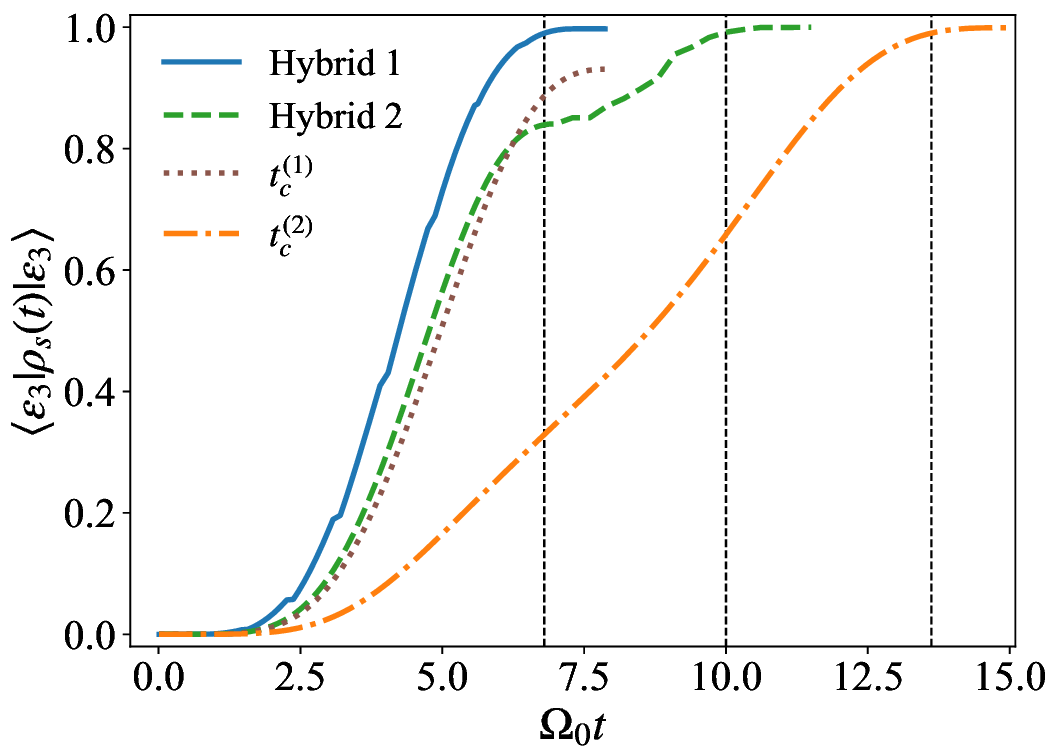}
\caption{Population on $|\varepsilon_3\rangle$ under various strategies. In Hybrid~1, the original control time for STIRAP is set as $t_c=7/\Omega_0$ and $M=10$ rounds of the state purification are discretely implemented during the integrated procedure. The full running time is then $t_c^{(1)}=t_c+\tau_p$, where $\tau_p=\sum_{i=1}^{10}\tau_i$ is the accumulated time for purification. In Hybrid~2, the control time $\tau_c$ is the same as that in Hybrid~1, after which $M=15$ rounds of measurements are repeatedly performed to achieve almost a unit fidelity as in Hybrid~1. The reference measurement intervals for the purification process in both Hybrid~1 and Hybrid~2 are set as $\tau_0=0.1/\Omega_0$ and $\tau_0=0.3/\Omega_0$, respectively. The coupling strength between the ancillary qubit and the system is $g_a=10\Omega_0$. The brown-dotted line labelled with $t_c^{(1)}$ indicates an unfaithful STIRAP within a control time equivalent to the full time for Hybrid~1. The orange dot-dashed line labelled with $t_c^{(2)}$ indicates a faithful STIRAP with a sufficiently long control time $t_c^{(2)}=15/\Omega_0$. The vertical black-dashed lines indicate the moments when the population on $|\varepsilon_3\rangle$ approaches $0.99$ under various strategies except the unfaithful STIRAP.}\label{Fig:STIRAP}
\end{figure}

To explore the purification-induced acceleration of the population transfer when the system deviates from adiabatic evolution, we propose two hybrid models combining STIRAP and measurement-based purification. During the stage of STIRAP, the system is driven by the Hamiltonian~(\ref{StirapHam}) with $\Omega_{12}(t)=\Omega_0f(t)$ and $\Omega_{23}(t)=\Omega_0[1-f(t)]$, where $\Omega_0$ represents the maximal magnitude of driving strength and $f(t)$ is a dimensionless function that satisfies $f(0)=0$ and $f(t_c)=1$ with a desired control time $t_c$. Here we take the hyperbolic sine function $f(t)=\sinh(ct/\tau)$, where $c$ is a scaling factor for the boundary conditions. During the state purification, the driving field between $|\varepsilon_2\rangle$ and $|\varepsilon_3\rangle$ is temporally switched off, i.e., $\Omega_{23}=0$, and the Rabi frequency of the driving field between $|\varepsilon_1\rangle$ and $|\varepsilon_2\rangle$ is set as a constant $\Omega_{12}=\Omega_0$. In the first hybrid model, the adiabatic evolutions and the purification processes present alternatively on stage. In particular, the original control time $t_c$ for both $\Omega_{12}(t)$ and $\Omega_{23}(t)$ can be divided into $M$ parts. A round of state purification consisting of a free evolution lasting a random $\tau_i$ and an instantaneous measurement on the ancillary qubit is performed at the end of each part. Taking account the time for $M$ rounds of state purification, the full running time for the first hybrid model is $t_c^{(1)}=t_c+\tau_p$, where $\tau_p=\sum_{i=1}^M\tau_i$. In the second hybrid model, the state purification is performed after the accelerated (diabatic and unfaithful) STIRAP is completed. In another word, the final state of STIRAP is the initial state for starting the purification process.

In Fig.~\ref{Fig:STIRAP}, we demonstrate the dynamics of the population on $|\varepsilon_3\rangle$ using the two hybrid models. To show the power of our state-purification scheme, we also present two results under the pure strategy of STIRAP. One is unfaithful with a shorter running period (see the brown-dotted line) and another one is faithful with a much longer period (see the orange dot-dashed line). The ``Hybrid~1'' strategy indicates that the STIRAP is divided to a certain number of parts and concatenated with discrete rounds of state purification. And ``Hybrid~2'' describes that an intact STIRAP is followed with the purification by measurements. It is found that the strategy of Hybrid~1 (see the blue solid line) prevails the other strategies in the running period for population transfer. In particular, ``Hybrid~1'', ``Hybrid~2'', and ``$t_c^{(2)}$'' cost $6.9/\Omega_0$, $10.0/\Omega_0$, and $13.6/\Omega_0$ in time to achieve $F=0.99$, respectively. With no assistance from the state purification, one can find that the same running period for the ``$t_c^{(1)}$'' strategy is too short to achieve a faithful population transfer. Both hybrid strategies overwhelm the faithful STIRAP of the ``$t_c^{(2)}$'' strategy and the success probabilities for ``Hybrid~1'' and ``Hybrid~2'' are respectively about $P_s=69\%$ and $P_s=84\%$.

\section{Discussion and Conclusion}\label{Sec:Conclusion}

Motivated by the inverse engineering or steering with directly performing a dense sequence of measurements on the interested system along a predesigned path~\cite{AharonovZeno,NonselectiveAharonovZeno}, our framework of POVM on the interested system by indirected measurements on the ancillary system provides a much broader regime for purification by measurements. It also suggests that local operations can be used to control a much larger coupled system. Previous works about the projection-based purification~\cite{MeasurementPurification,Nakazato2004,Nakazato2005} are devoted to optimizing the measurement intervals to enhance the population of the target state with a given interaction. They place a severe constraint over the target states and suffer from the purification inefficiency under a significant systematic error about the measurement intervals. Our scheme in this work, however, is mainly based on the purification operators and is capable of preparing an arbitrary eigenstate of the system with random time intervals. So that it is naturally robust against the errors with respect to the measurement moment and does not require a hybrid quantum-classical feedback control with observing the system state~\cite{QuantumClassiclHybrid,HardwareEfficient}. In sharp contrast to the steering protocol based on the nonselective measurements, our scheme does not involve with the complex many-body interaction and multiple ancillary qubits in preparing the Bell state.

In summary, we present an eigenstate purification framework by repeatedly measuring the ancillary qubit coupled to the system. The purification operator is built up without direct transitions towards the target state. Thus the measurements on the initial state of the ancillary qubit induce positive operator-valued measures that can purify the system into an arbitrarily chosen eigenstate by filtering out the populations over all the other states. In qubit systems, we apply our purification scheme to generate Bell states and GHZ states. Integrated with the conventional STIRAP protocol in a three-level system, we realize a much accelerated adiabatic population transfer with a high success probability. Our scheme can serve as a promising candidate for error correction when the system state deviates from the desired one due to the environment-induced decoherence, which is of great interest in the era of noisy intermediate-scale quantum. Much broadly, our scheme on quantum measurement contributes to a generic state preparation of the multi-particle system.

\section*{Acknowledgments}

We acknowledge financial support from the National Natural Science Foundation of China (Grant No. 11974311).

\bibliographystyle{apsrevlong}
\bibliography{ref}

\end{document}